\documentclass[wias]{iosart2c}     

\usepackage[T1]{fontenc}
\usepackage{times}%
\usepackage{graphicx} % for pdf, bitmapped graphics files
 \usepackage{balance}
 \usepackage{flushend}
\usepackage{verbatim}
\usepackage{algorithm}
\usepackage[noend]{algpseudocode}
\usepackage{float}
\usepackage{amsmath}
\usepackage{amssymb}
\usepackage[tight,footnotesize]{subfigure}
\pubyear{0000}
\volume{0}
\firstpage{1}
\lastpage{1}

\begin{document}

\begin{frontmatter}    

%\pretitle{}
\title{On-chip Face Recognition System Design with Memristive Hierarchical Temporal Memory}
\runningtitle{Memristive Hierarchical Temporal Memory}
%\subtitle{}
\author[]{Timur Ibrayev\fnms{} \snm{}},
\author[]{Ulan Myrzakhan\fnms{} \snm{}},
\author[]{Olga Krestinskaya\fnms{} \snm{}},
\author[]{Aidana Irmanova\fnms{} \snm{}},
\author[]{Alex Pappachen James\fnms{} \snm{}\thanks{Email:apj@ieee.org}}
\runningauthor{T. Ibrayev,U. Myrzakhan,O. Krestinskaya, A. Irmanova, A. P. James }
\address[]{School of Engineering, Nazarbayev University}

\maketitle

% For one author:

% Two or more authors:
%\author[A]{\fnms{} \snm{}\thanks{}},
%\author[B]{\fnms{} \snm{}}
%\runningauthor{}
%\address[A]{}
%\address[B]{}

\begin{abstract}
Hierarchical Temporal Memory is a new machine learning algorithm intended to mimic the working principle of neocortex, part of the human brain, which is responsible for learning, classification, and making predictions. Although many works illustrate its effectiveness as a software algorithm, hardware design for HTM remains an open research problem. Hence, this work proposes an architecture for HTM Spatial Pooler and Temporal Memory with learning mechanism, which creates a single image for each class based on important and unimportant features of all images in the training set. In turn, the reduction in the number of templates within database reduces the memory requirements and increases the processing speed. Moreover, face recognition analysis indicates that for a large number of training images, the proposed design provides higher accuracy results (83.5\%) compared to only Spatial Pooler design presented in the previous works.   
\end{abstract}


\end{frontmatter}

%%%%%%%%%%% The article body starts:

%\section{}\label{s1}

%\subsection{}\label{s1.1}

\section{Introduction}

The Hierarchical Temporal Memory (HTM) is a cognitive learning algorithm developed by Numenta Inc.~\cite{george2005neural}. HTM was designed based on various principles of neuroscience and, therefore, is said to be able to emulate the working principle of neocortex, a part of the human brain responsible for learning, classification, and making predictions~\cite{hawkins2006hierarchical}.
   
%The Hierarchical Temporal Memory (HTM) is a cognitive learning algorithm developed by Numenta Inc.~\cite{george2005neural}. By combining various principles of neuroscience, HTM is said to be able to emulate the working of the neocortex, the part of the human brain that is responsible for learning, classification, and making predictions~\cite{hawkins2006hierarchical}. 

%%%%%%%%%%%%%%%%%%  !!!!!!!!!!!!!!!!!!!! %%%%%%%%%%%%%%%%%%%%

After successful software realization of this learning algorithm, several works such as ~\cite{melis2009evaluation,zyarah2015design,fan2016Hierarchical,ibrayevdesign} have been conducted to realise its hardware implementation, including ~\cite{fedorova2016htm}, one of the latest works that propose analog circuit design of HTM Spatial Pooler based on the combination of memristive crossbar circuits with CMOS technology. One of the main advantages of the latest work is that the processing of input data is performed in the analog domain, which indeed can offer higher processing speed, primarily, due to the absence of analog-to-digital and digital-to-analog converters used in digital systems. Thus, inspired by design described in~\cite{fedorova2016htm} and from the idea of creating new analog add-on system that may move processing from digital domain to analog domain at the sensory level, this work proposes a system design of Hierarchical Temporal Memory for face recognition applications.

%There are few works done on the realization of hardware design for HTM~\cite{melis2009evaluation,zyarah2015design,fan2016Hierarchical,ibrayevdesign}, including one of the latest works~\cite{fedorova2016htm} that proposed analog circuit design of HTM Spatial Pooler based on a combination of memristive crossbar circuits and CMOS technology. Inspired from that design and from an idea of creating new analog add-on system that may move processing from digital domain to sensory level, this work proposes a system for Hierarchical Temporal Memory for face recognition application. 

%Despite the fact that there are few works done on the realization of hardware design for HTM, these works~\cite{melis2009evaluation,zyarah2015design,fan2016Hierarchical,ibrayevdesign,fedorova2016htm} illustrate that HTM has promising capabilities not only in feature extraction and pattern recognition, but also as learning add-on system. One of such recent works \cite{fedorova2016htm} proposed circuit design of HTM Spatial Pooler which is based on a combination of memristor devices and CMOS technology.

In particular, this work proposes a system level design for HTM that exploits the combination of a memristive crossbar based Spatial Pooler~\cite{fedorova2016htm}, and conceptual analog Temporal Memory based on learning mechanism inspired from the Hebbian rule. Further, we present face recognition algorithm for the proposed system and provide its performance results and analysis.

%\ Contribution: 1) System Level Design of HTM based on the combination of memristive crossbar based SP and hebbian rule based TM 2) Algorithm for the proposed system for face recognition applications 3) Verification of the proposed system on the face recognition applications

%Hence, this work proposes circuits for implementation of both HTM Spatial Pooler and Temporal Memory, which are built around \cite{fedorova2016htm}. The design is based on utilization of memristive crossbars for HTM Spatial Pooler implementation and incorporation of multi-leveled memory cells~\cite{mostafa2016process} for implementation of HTM Temporal Memory. In contrast to the previous design \cite{fedorova2016htm}, the proposed architecture introduces learning mechanism that offers advantages in terms of reduced memory requirements and processing time. Instead of saving multiple feature extracted images, HTM Temporal Memory produces single image combining features of all training images, resulting in fewer memory requirements. Moreover, the time required to perform pattern matching becomes less, as the number of feature extracted images reduces to a single image per class.

\section{Background}

\subsection{Hierarchical Temporal Memory}

The main core of the proposed system is HTM, that consists of two parts: Spatial Pooler and Temporal Memory \cite{hawkins2006hierarchical}. Spatial Pooler (SP) is responsible for the generation of the sparse distributed representations of input data and can be used for feature extraction and pattern recognition applications on its own, whereas Temporal Memory (TM) is responsible for learning input patterns and making predictions based on the temporal changes in the given input stream \cite{hawkins2010hierarchical}.

HTM was initially developed as the software algorithm \cite{hawkins2010hierarchical} and some research works, such as \cite{Farahmand2009Online, Ramli2015Pattern, Csapo2007Object}, were presented to illustrate and verify the capabilities of algorithmic implementation of HTM for performing classification, learning patterns, detecting abnormalities, and making predictions.  

Since HTM is a new machine learning algorithm, few attempts were made to implement it in hardware level. For instance,~\cite{melis2009evaluation} presented HTM hardware design for digital application-specific integrated circuit (ASIC) architecture,~\cite{zyarah2015design} depicts the design for the FPGA implementation of digital HTM,~\cite{fan2016Hierarchical} proposed computing blocks for HTM using memristive crossbar arrays and spin-neurons, which process data in both digital and analog domains, and the latest works~\cite{ibrayevdesign,fedorova2016htm} proposed circuits for HTM Spatial Pooler based on memristive crossbar architecture.

\subsection{Hebbian theory}
Introduced by Donald Hebb in 1949, Hebbian theory (also known as the Hebbian rule or Hebbian postulate) serves as one of the many learning mechanisms used in the design of artificial neural networks. In particular, it describes the basic idea behind synaptic plasticity and states that synaptic efficacy increases when a presynaptic cell takes part in repeated or persistent stimulations and firing of neighboring postsynaptic cell~\cite{fyfe2007hebbian}. Following the Hebbian rule, the activation of postsynaptic units in neural nets depends on the weighted activations of presynaptic units, which can be represented by (\ref{HR1})

\begin{equation}\label{HR1}
y_{i} = \sum_{j} w_{ij}x_{j}
\end{equation}

where $y_{i}$ represents the output of neuron $i$, $x_{j}$ stands for the $j^{th}$ input, and $w_{ij}$ is the weight of the connection from neuron $x_{j}$ to $y_{i}$ ~\cite{fyfe2007hebbian}. 

%In (\ref{my_first_eqn}) we have a very well known result.

In other words, the Hebbian theory claims that the synaptic weight between two neurons increases when both neurons simultaneously experience the activation or deactivation, and it decreases when they activate or deactivate separately. The equation for the change in synaptic weight $\Delta w_{ij}$ of the connection can be shown by (\ref{HR2}), which is known as a learning mechanism of Hebbian theory~\cite{fyfe2007hebbian}.  

\begin{equation}\label{HR2}
\Delta w_{ij} = \eta x_{j} y_{i}
\end{equation}

where $\eta$ is the learning rate. This learning mechanism in the artificial neural networks is used to alter weights between neurons.

\begin{figure*}[ht!]
\includegraphics[width=1\textwidth]{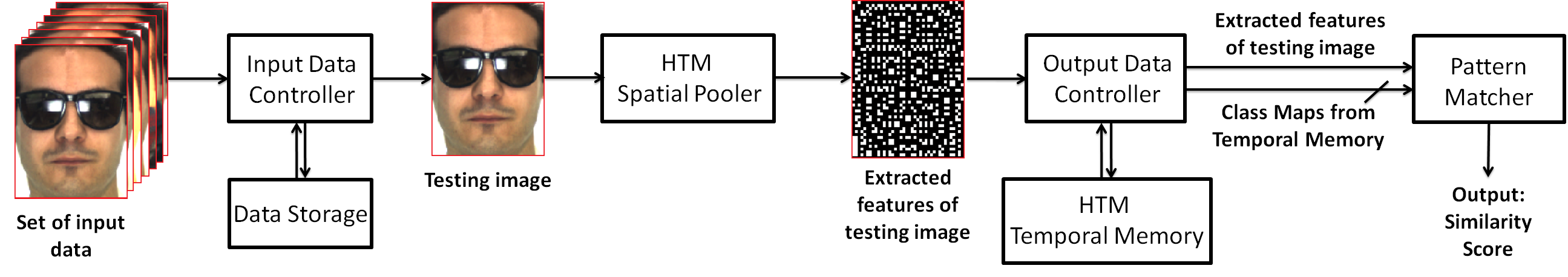}
\centering
\caption{High level block diagram of the proposed system illustrating operating principle of the HTM spatial pooler}
\label{SystemOverview}
\end{figure*}

% \subsection{Memristor}
% Development of machine learning algorithms based on neuroscientific principles, such as neurons and synaptic connections, demanded to move from traditional CMOS only designs to modern technologies. One of such innovations that has already shown its practical application in the implementation of large scale learning systems is a memristor device \cite{kozma2012advances,maan2016survey}.

% Manufactured for the first time in 2003 by HP Labs, memristor device, unlike the other technologies, has a distinctive ability to change its state with respect not only to the current input but also to the history of the previous inputs \cite{strukov2008missing}. This capability of the device makes it invaluable in an attempt to recreate main neuronal functions, such as memorization and data processing. Moreover, memristor gives advantages in terms of lower on-chip area and small power dissipation associated with absence of leakage currents, in contrast to semiconductor resistors and CMOS technology \cite{strukov2008missing}.

\subsection{Memristor models}
Currently, there are several available memristor models, which incorporate not only the characteristics of real existing devices, but also provide the possibility to switch from one device parameters to another, so that suitability of these devices can also be assessed. For example, \cite{yakopcic2013memristor} proposed a model with nanosecond switching time, which is crucial in designing real-time systems. Recent models proposed in~\cite{biolek2016} allow simulation of large-scale networks of memristors as well since parallelism and scalability of the system play important role in processing huge amount of data.

\section{Proposed HTM System}
\subsection{System Design} 
%%%%%%%%%%%%%%%%%%  !!!!!!!!!!!!!!!!!!!! %%%%%%%%%%%%%%%%%%%%
A high-level block diagram of the proposed system is illustrated in Fig.~\ref{SystemOverview}. Input data controller reads the input image, places it in data storage and sends this input image to HTM Spatial Pooler by partially retrieving it from the data storage. The data controller is substantial as partial sending is required to ensure that the selected size of HTM SP is capable of processing an entire image. In turn, HTM SP is responsible for feature extraction of the input image and thus, provides a binary output. If the input image is a training image, the output data controller directs its extracted features to HTM TM, which creates a single image/template for each class having common features of all the training images belonging to that particular class. During the testing phase, resulting images stored within TM are then used by pattern matcher to calculate the similarity score between the input testing image and each of the trained classes.

\subsection{System Algorithm}

\begin{algorithm}[t]
%\label{algg1}
\caption{for the proposed HTM face recognition system}\label{alg:matlab}
\begin{algorithmic}[1]

%PREProS 

\\
\Comment{\textbf{Pre-processing of input images}}
\State $Convert\; RGB\; image \; into \;grayscale$
\State $Filter\; grayscale\; image\; with\; standard\; deviation\; filter$
%HTMSP
\\
\Comment{\textbf{HTM SP}}
\State $Create\; random\; matrix\; w\; of\; N\times N \;size$

 \ForAll{$n\; in\; w$}
 \If{$w(n) \geqslant \gamma$} 
 %\Comment{$where$ $element$ $w_i \in w$}
 \State $w(n) = 1$
 \Else
 \State $w(n) = 0$
 \EndIf
 \EndFor
\State$Divide\;image\;into\;blocks\;of\;N\times N\; pixels$
\For {$m$ image blocks}  
\State $column.overlap(m) = sum (w\times image.block(m))$ 
\EndFor
\State$Divide\;image\;into \; inhibition \;blocks\;  of\; M\;\times M\; columns$
\For{$c$ columns \textbf{within} \textit{inhibition.block($M\times M$)}}
\State$\theta = max (column.overlap(c.columns))$
\If{$column.overlap(c) \geq \theta$} 
%\Comment{$where$ $element$ $w_i \in w$}
\State $inhibition.block(c) = 1$
\Else
\State $inhibition.block(c) = 0$
\EndIf
\EndFor
\State $SP.image = inhibition.block$
\\
\Comment{\textbf{HTM TM}}
\If{training stage}
\State$Define\;number\;of\;train\;classes\;k$
\State$Define\;number\;of\;train\;images\;z\;per\;class$
\ForAll {$n$ in $k$ train classes}

\ForAll{$j$ in $z$ train images}
\ForAll{$i$ pixels in $SP.image$}
\If {$SP.image(j,i)$ = 1}
\State$\begin{aligned}[]
             class.map(n,i)=class.map(n,i) +\Delta
         \end{aligned}$
\ElsIf {$SP.image(j,i)$ = 0}
\State$\begin{aligned}[]
             class.map(n,i)=class.map(n,i) -\Delta
         \end{aligned}$
\EndIf
\EndFor
\EndFor
\ForAll{$i$ pixels \textbf{in} $class.map(n)$}
\If {$class.map(n,i) \geq \sigma$}
\State$class.map(n,i) = 1$
\ElsIf {$class.map(n,i) < \sigma$}
\State$class.map(n,i) = 0$
\EndIf

\EndFor
\EndFor
\\
\Comment{\textbf{Recognition stage and image classification}}
\ElsIf{testing stage}

\EndIf
\ForAll{$n$ in $k$ image classes}
\For{$i$ image pixels}
\State$xor.res(n) = XOR(class.map(n,i),SP.image(i))$
\State$score(n)=sum(xor.res(n))$
\EndFor
\EndFor
\State$determined.class=class(min(score))$
\end{algorithmic}
\end{algorithm}

In this work, we also propose Algorithm~\ref{alg:matlab} that can be used to analyze the effectiveness of the proposed system. The algorithm shows interconnections between main processing stages of the entire system: pre-processing, HTM SP, HTM TM, and pattern matching. Pre-processing stage, shown in lines 2-3 in Algorithm~\ref{alg:matlab}, is necessary to convert the input image into system compatible format. In this stage, we convert input image to grayscale and enhance its quality using standard deviation filtering. This is achieved by either external means or by input data controller.

HTM SP stage models feature extraction process achieved by the Spatial Pooler block that is illustrated in lines 5-21 of Algorithm \ref{alg:matlab}. This is done by initially generating random weights matrix $w$, so that each weight in $w$ would have analog value between $0$ and $1$. The weights matrix has dimensions of $N\times N$, which also defines dimensions of each column within Spatial Pooler. Lines $6$-$10$ define connectivity of each synapse, so that if its weight $w$ is higher than the threshold $\gamma$, the synapse is connected and represented by $1$,  but if $w < \gamma$, it is disconnected and represented by $0$. Synapse connectivity is used to determine overlap value $column.overlap()$ for each column $m$, which is represented as the $sum$ of the products of synaptic weight matrix $w$ and $N\times N$ bits of the image within the region $m$. This overlap value represents the importance of bits connected to each particular column.

Lines $15$-$20$ define the inhibition rule implemented in the proposed system. According to the rule, inhibition is performed in a block-by-block manner, each having dimensions of $M\times M$ columns and is based on overlap values achieved by columns $c$ lying within that inhibition block $inhibition.block()$. This is done by comparing individual overlap values of columns $c$ with threshold value $\theta$, which is determined as the maximum overlap that is detected within that particular inhibition block. Then, the column or columns with overlap value greater than or equal to $\theta$ are considered as important and represented by logical high $1$ value. Otherwise, columns are considered as unimportant and represented by logical low $0$ value. As a result, the binary feature extracted output image $SP.image$ after HTM SP processing is formed by concatenating all inhibition blocks.

HTM TM stage defines a learning mechanism that is activated during training stage when binary feature extracted an image from HTM SP $SP.image$ is moved by the output data controller block to the HTM TM block. Lines $26$-$32$ define that proposed TM should create certain $class.map$ for each class in $k$ image classes, which reflects temporal variations of spatial features. This is done by making TM update $class.map$ every time new feature extracted image is fetched to TM block. Based on whether bit has the value of $1$ or $0$ within $SP.image$, corresponding memory cell within $class.map$ is either increased or decreased by $\delta$ value, respectively. At the end of training phase, according to lines $33$-$37$, $class.map$ of each class is binarized.

Recognition and image classification stage defines a pattern matching process that is active during testing phase when binary feature extracted image $SP.image$ is moved by the output data controller block to the Pattern Matcher block. According to lines $39$-$44$, the similarity score between extracted features $SP.image$ of the testing image and each of the class maps stored within HTM TM is defined as the sum of XOR logic high $1$ outputs. Since XOR operation produces logic high $1$ output at places where two compared bits are of different value, a class of the tested image can be defined as the $class.map$ that produces the least $score()$ value.

\section{Circuits for the Proposed System}
\subsection{Spatial Pooler}

\begin{figure}[ht!]
\includegraphics[width=0.47\textwidth]{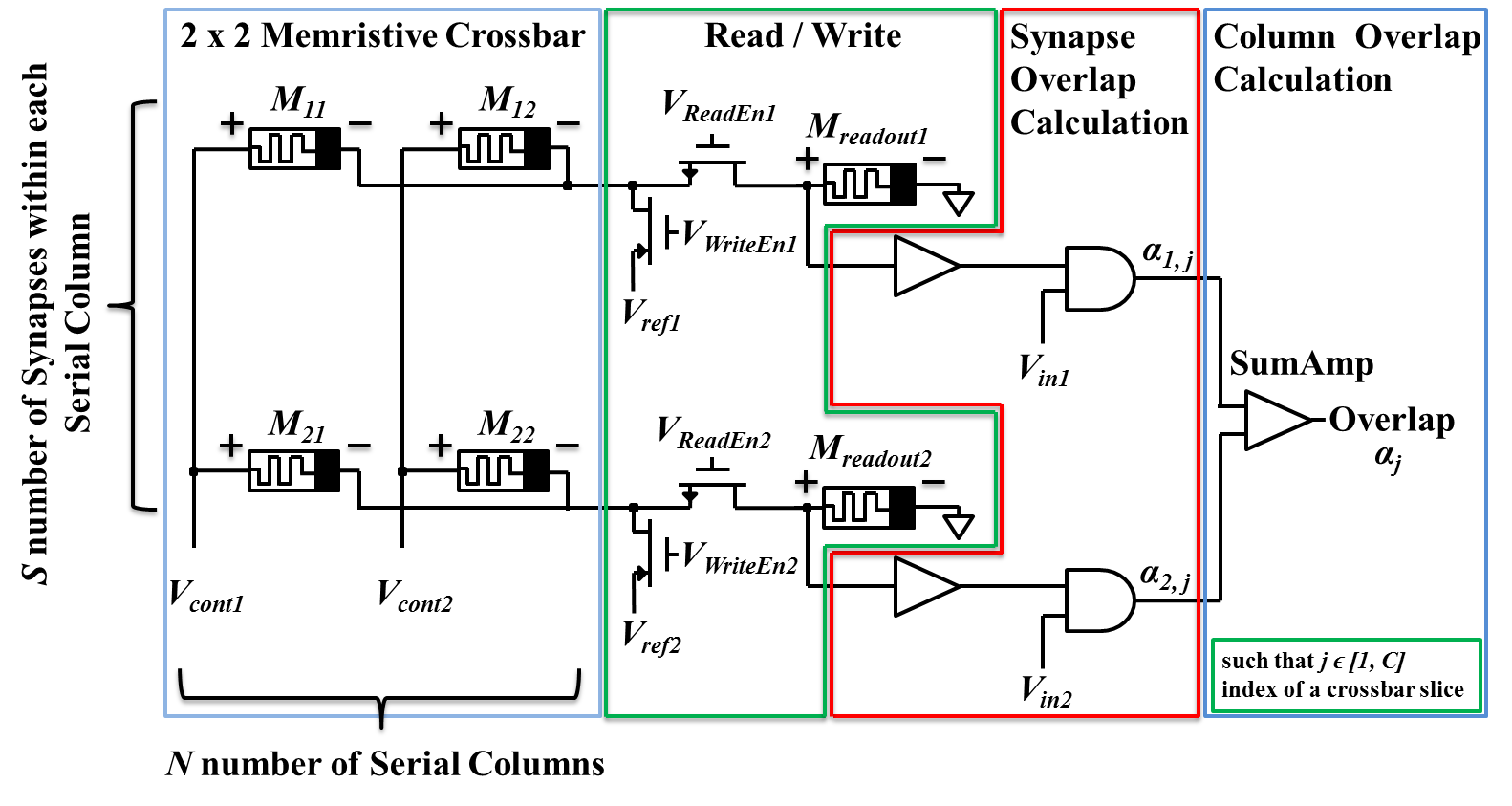}
\centering
\caption{Single memristive crossbar processing unit as presented in \cite{ibrayevdesign}}
\label{crossbarSP}
\end{figure}

\begin{figure}[ht!]
\includegraphics[width=0.47\textwidth]{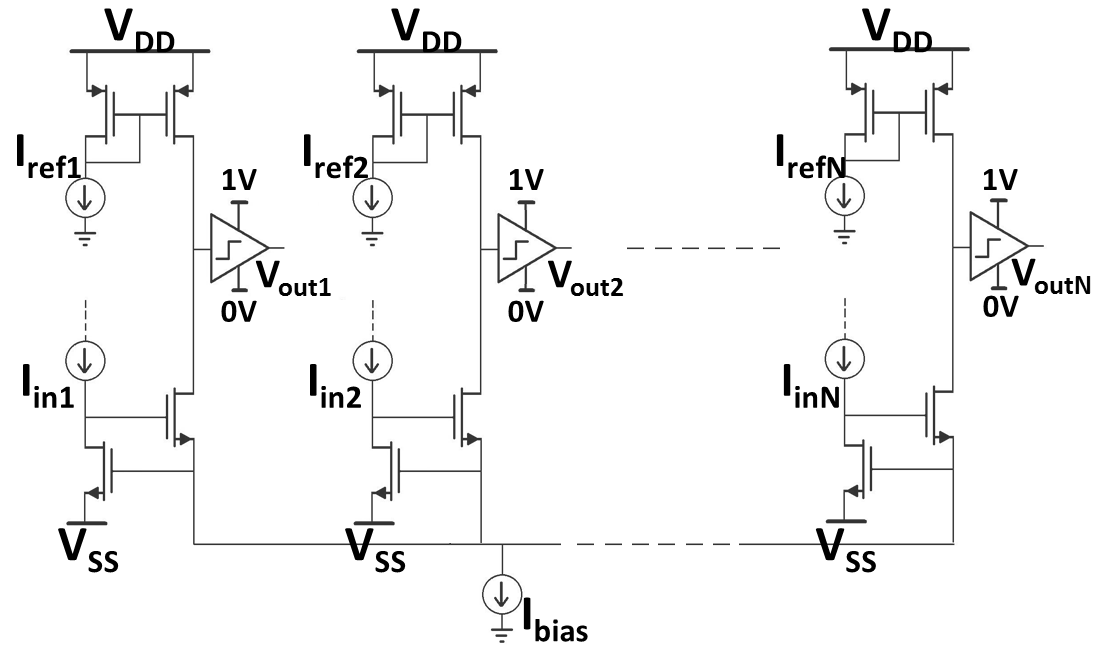}
\centering
\caption{Winner-Take-All Circuit taken from \cite{ibrayevdesign}}
\label{WTA}
\end{figure}

In the proposed system HTM SP processing can be implemented with memristive crossbar based SP~\cite{fedorova2016htm}. Memristor devices due to its ability to memories and being able to mimic neurons find various applications \cite{maan2016survey, maan2015memristor, olga2017htm}. Figure \ref{crossbarSP} illustrates single memristive crossbar processing unit. Memristive-CMOS circuits allow precise realization of the feature extraction process described by algorithm lines $3$-$10$ with additional advantages in terms of parallel synaptic processing and compact storage of synaptic weights. Figure~\ref{WTA} illustrates modified version of the WTA circuit designed by \cite{lazzaro1988winner}, which is when combined with SP circuits presented in~\cite{fedorova2016htm} allow implementation of inhibition processing described by algorithm lines $11$-$17$.

%The operating principle of the HTM Spatial Pooler is illustrated in Fig.~\ref{SystemOverview}. HTM Spatial Pooler is responsible for feature extraction of every input image and provides binary output i.e. important features of the image are represented by binary 1 and non-important features by binary 0.

%The circuit implementation of a single crossbar slice having $N = 2$ number of serial columns and $S = 2$ number of parallel synapses within each of the serial columns is shown in HTM spatial pooler block in Fig.~\ref{SystemOverview}. Each of the crossbar slices is connected to Winner-Take-All (WTA) circuit. Figure~\ref{WTA} illustrates modified version of the WTA circuit designed by \cite{lazzaro1988winner}. The main function of this circuit is to set the important features of the image to binary 1 and all the other features to binary 0. 

\begin{comment}
\begin{figure}[ht!]
\includegraphics[width=7cm]{spatial_pooler_TI_UM_2}
\centering
\caption{Block diagram illustration of a single region of the HTM hierarchy displaying input data, crossbar slices, and Winner-Take-All (WTA) circuit for inhibition phase implementation}
\label{SpatialPooler}
\end{figure}

\begin{figure}[ht!]
\includegraphics[width=7cm]{crossbar_edited_TI_UM}
\centering
\caption{Crossbar slice circuitry with N = 2 number of serial columns and S = 2 number of parallel synapses per serial column}
\label{CrossbarCircuit}
\end{figure}

\begin{figure}[!ht]
    \centering
    \subfigure[]
    	{
    	\includegraphics[width=7cm]{spatial_pooler_TI_UM_2}
    	\label{SpatialPoolerBlock}
		}

	\subfigure[]
		{
    	\includegraphics[width=6.5cm]{crossbar_edited_TI_UM}
    	\label{CrossbarCircuit}
		}
    \label{SpatialPooler}
    \caption{(a) Block diagram illustration of a single region of the HTM hierarchy displaying input data, crossbar slices, and Winner-Take-All (WTA) circuit for inhibition phase implementation. (b) Crossbar slice circuitry with N = 2 number of serial columns and S = 2 number of parallel synapses per serial column}
\end{figure}
\end{comment}

\subsection{Temporal Memory}

Instead of saving all the feature extracted images as it was done in the previous SP design \cite{fedorova2016htm}, the proposed work incorporates conceptual analog TM into the entire system. It is, in turn, intended to reduce memory requirements and processing time by creating single training image, which is called a \textit{\textbf{class map}} in this work. Such class map incorporates features of all training images belonging to a single class and allows pattern matching to be performed by comparing the testing image with the only single image for each of the memorized (trained) classes.

This is realized by making TM learn by \textit{focusing} on both important and unimportant features and by \textit{reflecting} how features change with time. 
%The architecture of Temporal Memory is based mainly on the learning mechanism of Hebbian theory given by .

\textit{Focus} is achieved by placing Temporal Memory circuitry after Spatial Pooler so that the inputs are not the original training images, but feature extracted images provided at the output of Spatial Pooler. Since each of these outputs is binary in nature, such placement allows Temporal Memory to differentiate important and unimportant features.

\textit{Reflection} is achieved by changing the weights of the Temporal Memory cells according to the importance of the corresponding input bit. This is realized by implementing learning mechanism of the Hebbian theory given by (\ref{HR2}), which, in general, is used to determine the weight change between presynaptic and postsynaptic units. In the proposed design of Temporal Memory, the binary pixels of feature extracted images are used as presynaptic units, whereas postsynaptic units are represented by a matrix of ones of the same size as the image. The realization of postsynaptic units as a matrix of ones is required to ensure that every pixel of feature extracted image is treated as equally important. Moreover, such arrangement allows alteration of the weights of Temporal Memory cells with respect to their importance. 

In particular, if the input bit of feature extracted image is $1$, meaning that it represents an important feature, then the weight of the corresponding Temporal Memory cell increases by \textit{positive weight update} (\(+\Delta\)) value. Contrary, if the input bit of the feature extracted image is $0$, meaning that it represents an unimportant feature, then the weight of the corresponding Temporal Memory cell decreases by \textit{negative weight update} (\(-\Delta\)) value. This algorithm of differentiating important and unimportant features using learning mechanism of the Hebbian rule is illustrated in Figure 4.  

\begin{figure}[ht!]
\includegraphics[width=8cm]{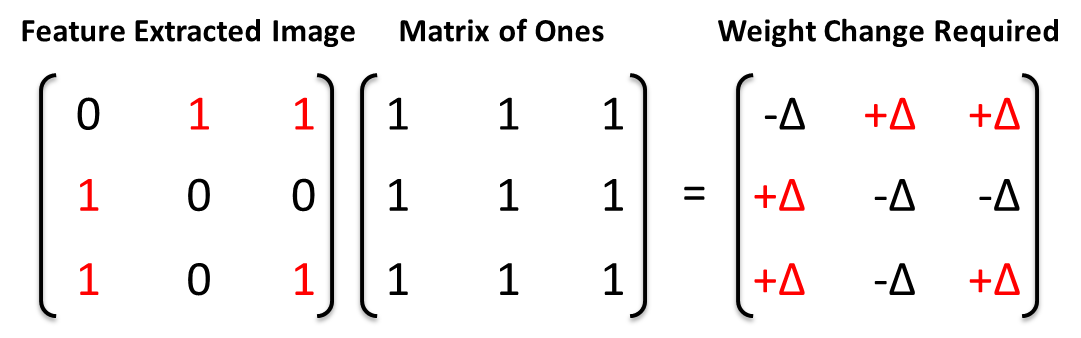}
\centering
\caption{Example of determining required weight updates (positive or negative) using Hebbian learning mechanism at each particular pixel within Temporal Memory}
\label{HR_Matrix}
\end{figure}

\begin{figure}[ht!]
\includegraphics[width=6cm]{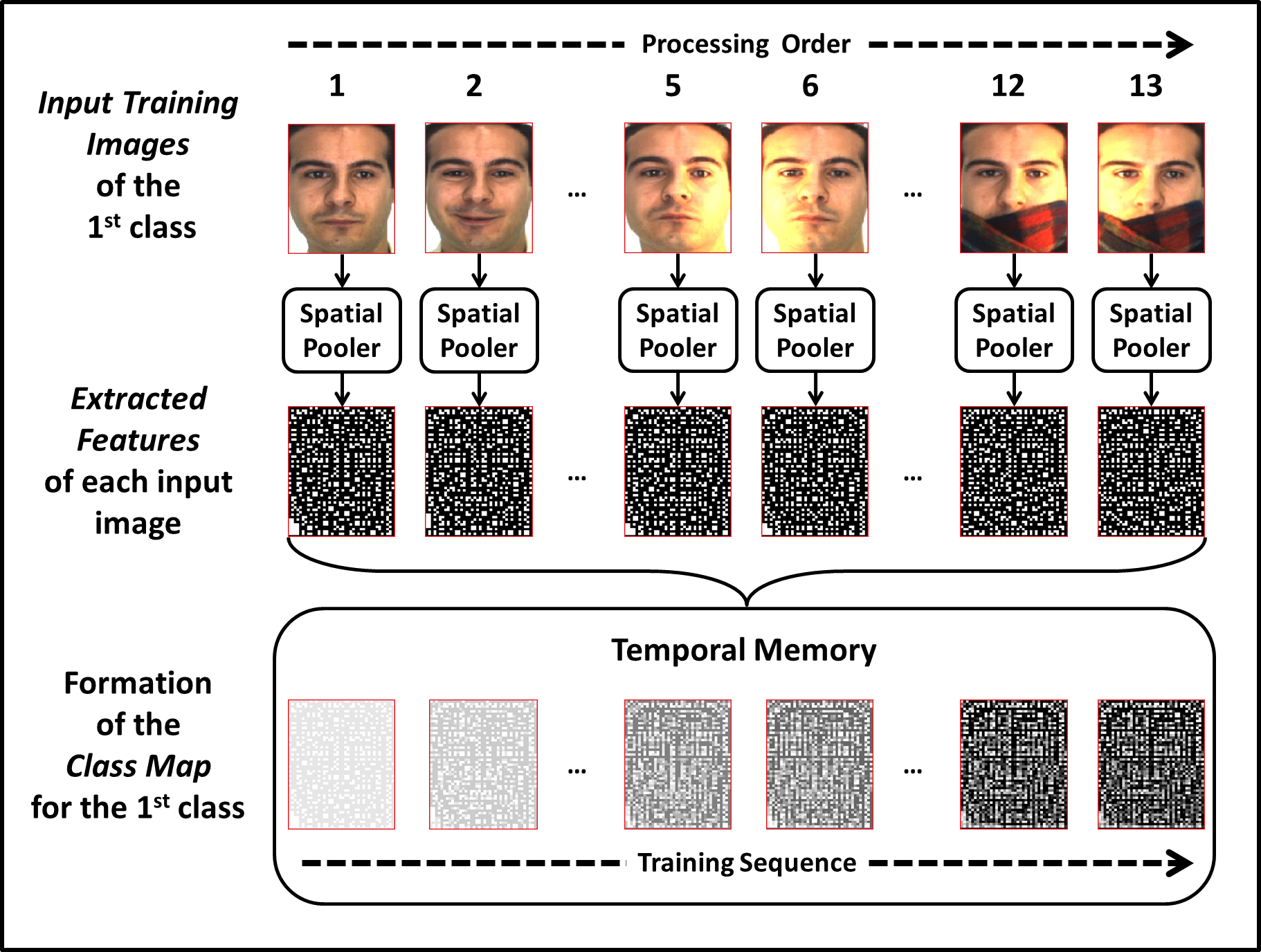}
\centering
\caption{The main principle of single class map formation using Temporal Memory and feature extracted images obtained from Spatial Pooler}
\label{Process}
\end{figure}

As a result, instead of having multiple binary images with extracted features, TM creates single analog image incorporating important and unimportant features of and having the same dimensions as each of the input images belonging to a single class. Figure~\ref{Process} illustrates the formation of the class map for the first class by fetching feature extracted binary images belonging to the first class to TM. All of the TM cells, initially having the same weight, eventually become distinguishable at the end of the training sequence.

However, such learning mechanism requires TM to be multi-valued. This is to ensure that the weights can take values not only of 1 and 0, as feature extracted images do, but can be changed according to the weight update value, which is \(\pm \Delta\). 

\begin{figure}[ht!]
\includegraphics[width=5cm]{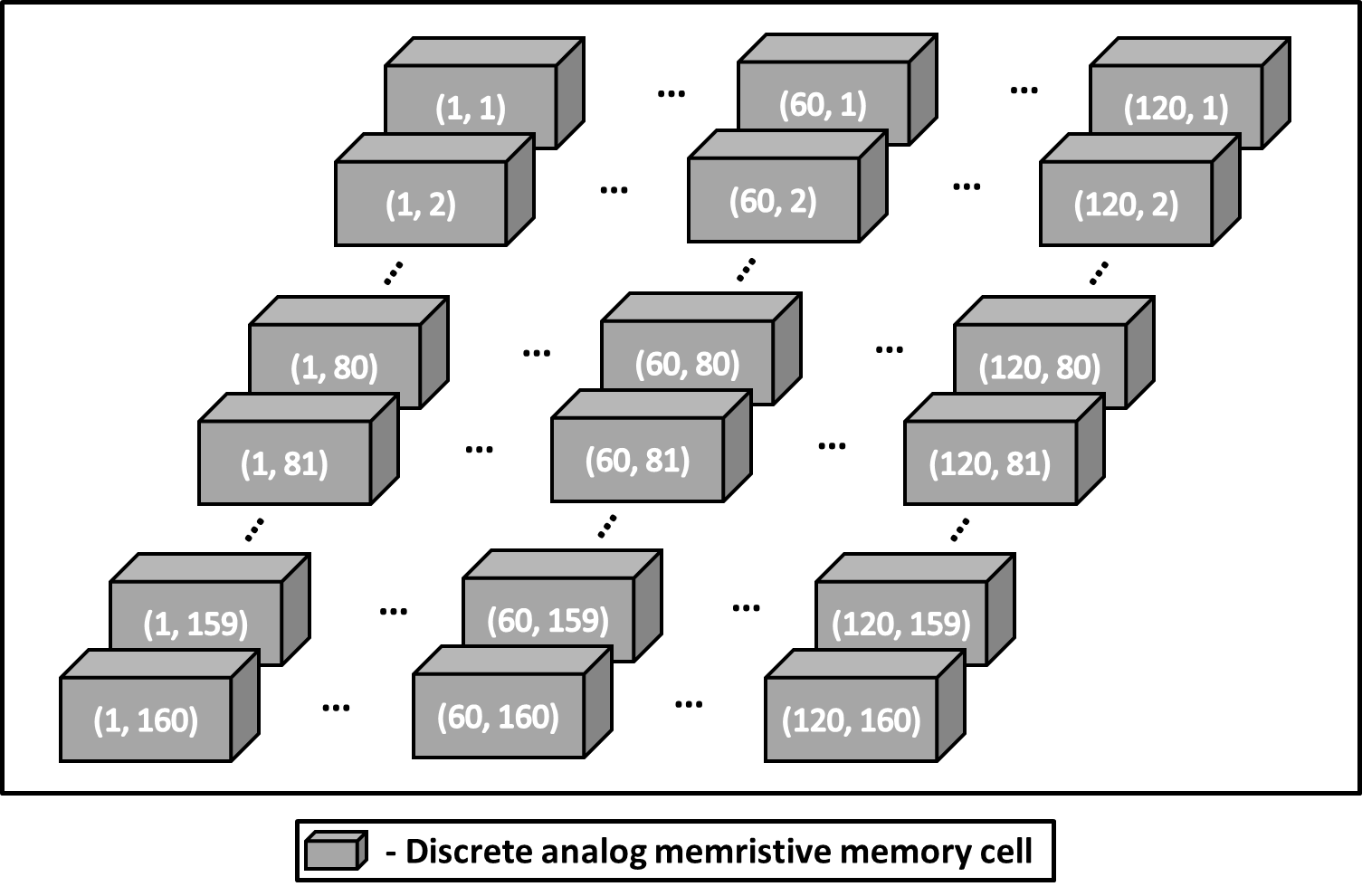}
\centering
\caption{The design of Temporal Memory consisting of \(120\times 160 = 19200\) memory cells and that is used for storing a single class map trained by fetching input images having dimensions of \(120~bits\times 160~bits\).}
\label{TempMemory_Blocks}
\end{figure}

Hence, Fig. \ref{TempMemory_Blocks} illustrates the design of TM required to memorize single class map and utilizing multi-valued memory cells. The total number of the required memory cells correspond to the product of the number of memorized classes and the number of bits of a single class map. For example, for 13 class maps, each having dimensions of \(120~bits\times 160~bits\), the required number of memory cells is \(13\times 19200\). The multi-valued memory cells, in turn, can be realized by using n-bit memristor-based memory, which is described in \cite{mostafa2016process}.

\subsection{Pattern Matcher}

\begin{figure}[!ht]
    \centering
    \label{XORgate}
    \subfigure[]
    	{
    	\includegraphics[width=0.18\textwidth]{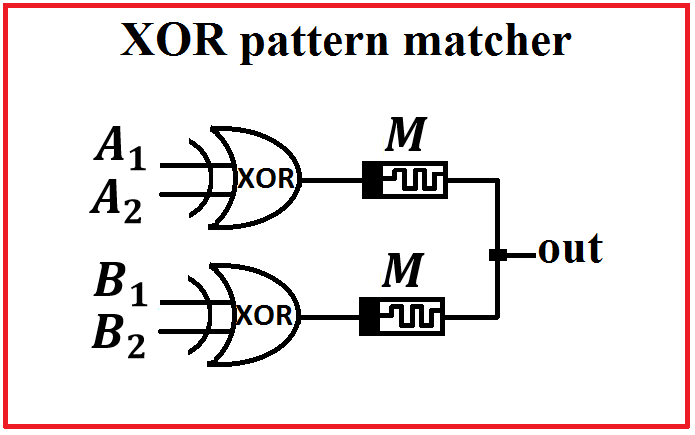}
    	\label{XOR:pattern matcher}
		}
	\subfigure[]
		{
    	\includegraphics[width=0.18\textwidth]{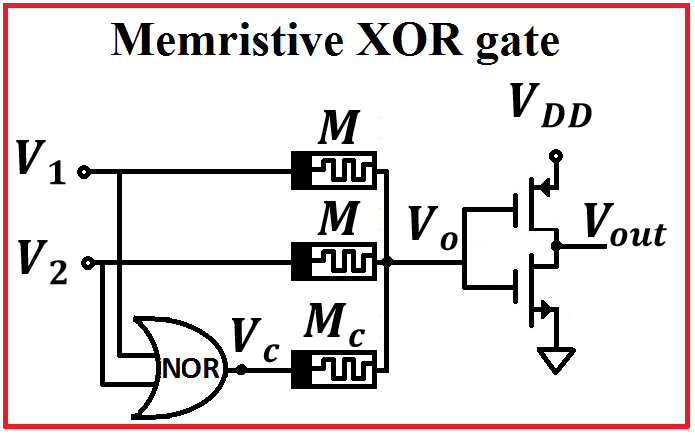}
    	\label{XOR:memristive}
		}
    \caption{(a) A 2 bit XOR pattern matcher, (b) XOR configuration of memristive memory threshold logic}
\end{figure}

After class maps were formed within TM during the training phase, the testing is performed by comparing each input testing image with all the class maps learned by the system. In the proposed system this can be achieved by fetching a feature extracted input testing image and all the class maps into memristive pattern matcher (Fig. \ref{XOR:pattern matcher}), which is realized by memristive XOR gates illustrated in Fig. \ref{XOR:memristive} and described in \cite{fedorova2016htm}. 

\begin{figure}[ht!]
\includegraphics[width=6cm]{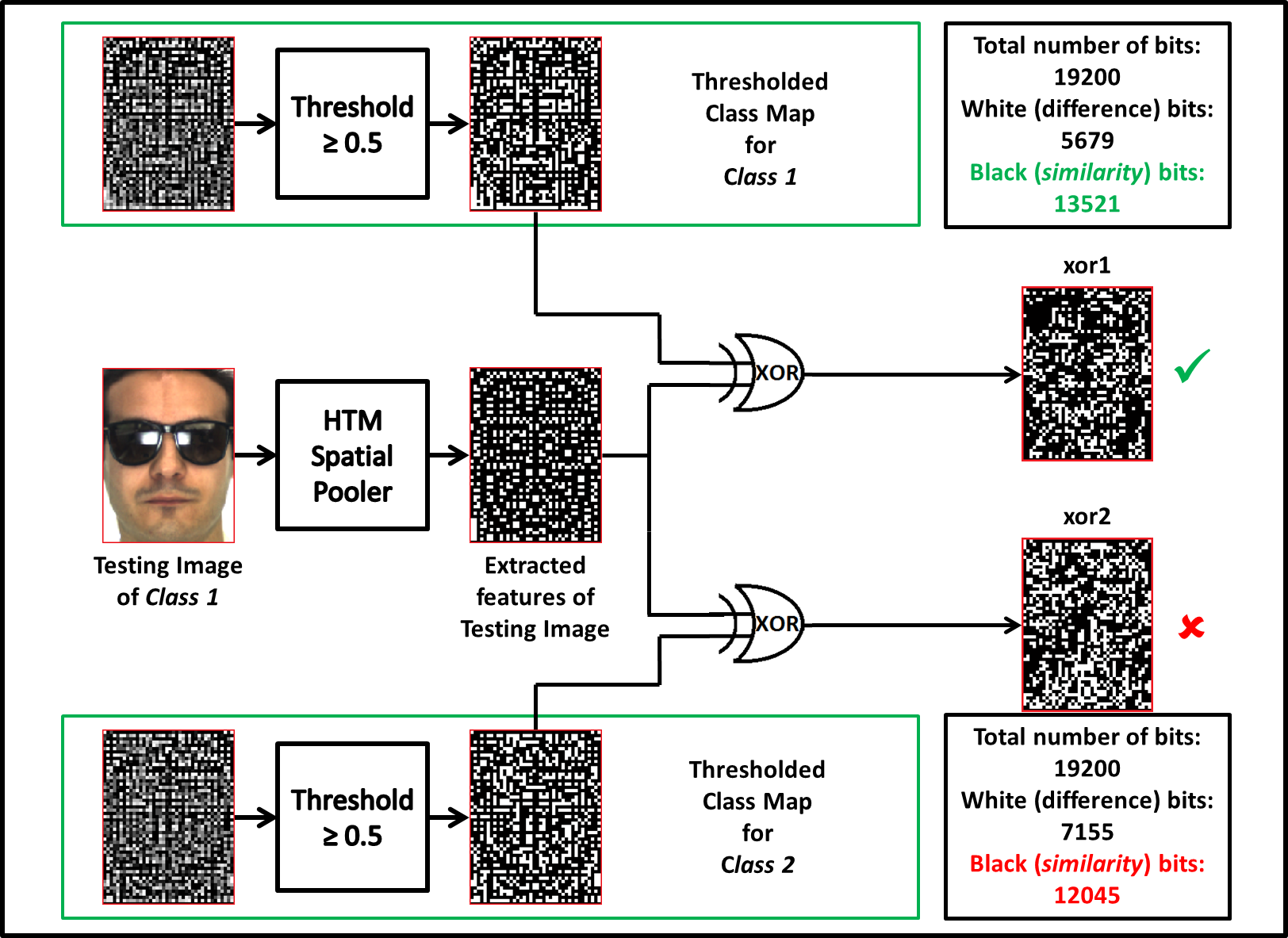}
\centering
\caption{The face recognition process done by calculating similarity score between the feature extracted testing image and two class maps using memristive XOR gates}
\label{Matching}
\end{figure}

Figure \ref{Matching} illustrates the principle used to determine a similarity score between any feature extracted the image and two arbitrary class maps. The class maps are thresholded at the mean value of $0.5$, so that XOR logic can be accomplished. For two input images, memristive XOR gates will produce output image having logical $0$ at the regions where both images represent important or unimportant features (i.e. either both have $1$ at that region or both have $0$ at that region) and having logical $1$ at the regions where two images represent different features. Hence, the class of an input testing image is determined by the class map that has the least number of white bits (or the greatest number of black bits) at the corresponding XOR output. 

The described pattern matching process emphasizes the advantage of the proposed design in terms of faster processing speed: the time required for the system to determine similarity score reduces, as the number of templates required to be compared with the input testing image decreases to a single image (that is \textbf{\textit{class map}}) per class.

\section{Results and Discussion}

The performance metrics were based on face recognition accuracy of AR database \cite{martinez1998ar}. The database consists of 100 classes each having 26 images that were taken in two sessions. These images then were divided into two separate sets. The first set consisted of 13 images of each class taken during the first session and was used to train Temporal Memory, whereas the second set consisted of 13 images of each class taken during the second session and was used for testing.

\begin{figure}[ht!]
\includegraphics[width=0.47\textwidth]{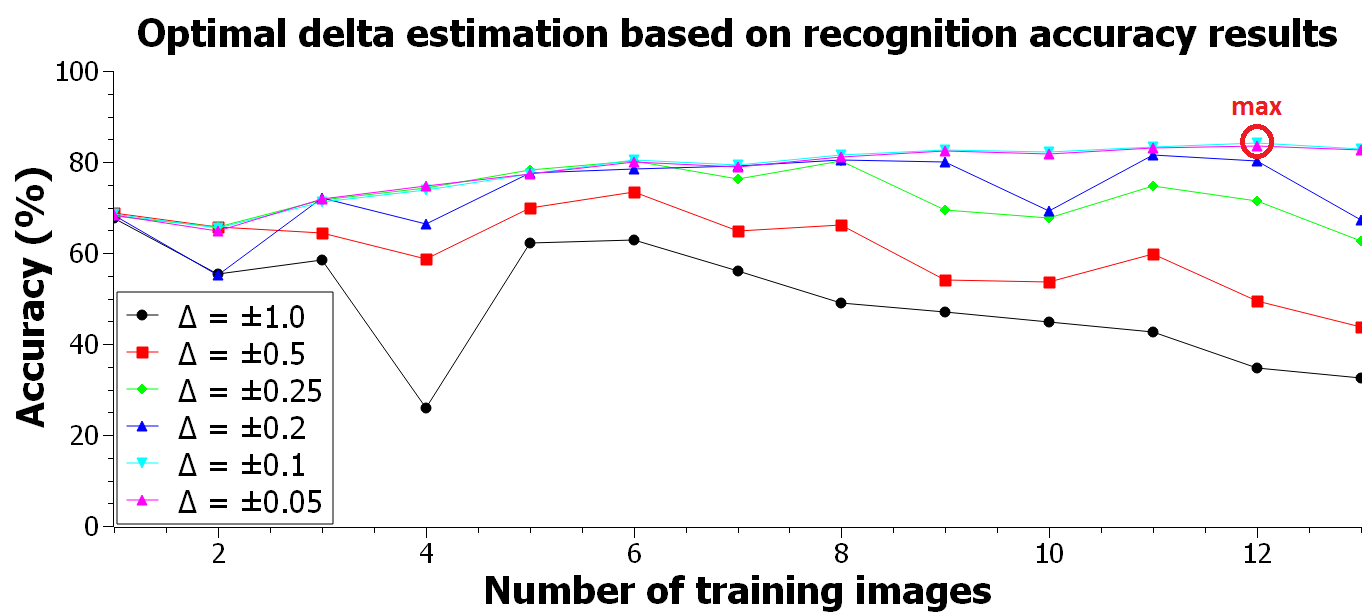}
\centering
\caption{Optimal Delta estimation based on recognition accuracy results}
\label{Delta}
\end{figure}

Based on the above mentioned set up, the first analysis was aimed to determine optimal delta (\(\pm \Delta\)) required to update the weights of Temporal Memory for a different number of training images. Figure~\ref{Delta} illustrates the recognition accuracy results achieved for different combinations of \(\pm \Delta\) and the size of the training set. As can be seen, for the number of training images between 1 and 13 the maximum recognition accuracy is achieved when \(\pm \Delta\) value is lower than or equal to $\pm 0.1$. Moreover, as the size of the training set increases, the maximum achieved accuracy increases for \(\pm \Delta\) value of $\pm 0.01$ and decreases for large values of \(\Delta\).

This result indicates that achieving maximum recognition accuracy with a large number of training images is possible, when the weight update value is small. Another point that should be taken into account is that the value of memristor directly proportional to the duration of applied constant voltage. These two statements imply that the increased number of training images decreases the duration applied voltage required to update the weights, which means the consecutive input images can be processed at a higher speed.

\begin{table*}[ht]
\centering
\renewcommand{\arraystretch}{1.3}
\caption{Recognition accuracy of classifying test images in each category of AR database done by two different architectures using single template or class map per each class.}
%\begin{threeparttable}
\label{Accuracy}
%\noindent\adjustbox{max width=\columnwidth}{%
\begin{tabular}{|p{4cm}|p{1.5cm}|p{1.5cm}|p{1.5cm}|p{1.5cm}|p{1.5cm}|}
\hline
Architecture & Emotions &  Light conditions  &  Occlusions  (glasses) &  Occlusions  (scarf)  & \textbf{Total} %\tabularnewline
%\hline
\\\hline
Spatial Pooler~\cite{fedorova2016htm} & $77.50\%$ & $91.00\%$ & $84.33\%$ & $53.33\%$ & $\textbf{76.54\%}$ %\tabularnewline
\\\hline
 Spatial Pooler and  Temporal Memory  & $84.25\%$ & $96.33\%$ & $85.67\%$ & $67.67\%$ & $\textbf{83.48\%}$ \\
\hline
\end{tabular}%

\end{table*}

After optimal weight update value was determined to be \(\pm 0.01\), the analysis was performed to compare the effectiveness of the architecture based on Spatial Pooler only~\cite{fedorova2016htm} and the proposed architecture combining Spatial Pooler and Temporal Memory on face recognition task. To make common settings, for the architecture of only Spatial Pooler the training images belonging to a single class were initially averaged and the averaged images then were processed by Spatial Pooler to provide a feature extracted training templates. For the proposed architecture, the training images were processed by Spatial Pooler and extracted feature outputs were used to create class maps within Temporal Memory. Table \ref{Accuracy} illustrates recognition accuracy results for the condition when both architectures had one template or class map for each of 100 classes (giving in total 100 templates or class maps) with which all of 13 testing images of each class (giving in total 1300 testing images) were compared. Comparing these results with these reported in~\cite{fedorova2016htm}, it can be seen that as the number of training images increases, the architecture incorporating Temporal Memory provides higher recognition accuracy at lower memory requirements and faster processing at pattern matching stage.

\section{Conclusion}

In this paper, we proposed system realization of HTM Spatial Pooler and Temporal Memory using memristor-CMOS circuits. The main difference from the existing HTM system with memristor is in that the system incorporates Temporal Memory with learning capability. The learning process of the system involves collecting important features from the training data of given class and creating its class map - a single image, based on the extracted features. Hence, the main advantage of the system is less memory occupation of HTM that provides with higher processing speed. The results of performance analysis indicate that for a large set of training images the proposed recognition system provides a higher accuracy compared to the results presented in the previous work.


%%%%%%%%%%%%%%%%%%%%%%%%%%%%%%%%%%%%%%%%%%%%%%%%%%%%%%%%%%%%%%%%%%%%%%%%%%%%%%%%

%%%%%%%%%%%%%%%%%%%%%%%%%%%%%%%%%%%%%%%%%%%%%%%%%%%%%%%%%%%%%%%%%%%%%%%%%%%%%%%%

%%%%%%%%%%%%%%%%%%%%%%%%%%%%%%%%%%%%%%%%%%%%%%%%%%%%%%%%%%%%%%%%%%%%%%%%%%%%%%%%
%\section*{APPENDIX}


%\section*{ACKNOWLEDGMENT}

%%%%%%%%%%%%%%%%%%%%%%%%%%%%%%%%%%%%%%%%%%%%%%%%%%%%%%%%%%%%%%%%%%%%%%%%%%%%%%%%

\bibliographystyle{IEEEtran}
\bibliography{main_reference}

\begin{thebibliography}{10}
\providecommand{\url}[1]{#1}
\csname url@samestyle\endcsname
\providecommand{\newblock}{\relax}
\providecommand{\bibinfo}[2]{#2}
\providecommand{\BIBentrySTDinterwordspacing}{\spaceskip=0pt\relax}
\providecommand{\BIBentryALTinterwordstretchfactor}{4}
\providecommand{\BIBentryALTinterwordspacing}{\spaceskip=\fontdimen2\font plus
\BIBentryALTinterwordstretchfactor\fontdimen3\font minus
  \fontdimen4\font\relax}
\providecommand{\BIBforeignlanguage}[2]{{%
\expandafter\ifx\csname l@#1\endcsname\relax
\typeout{** WARNING: IEEEtran.bst: No hyphenation pattern has been}%
\typeout{** loaded for the language `#1'. Using the pattern for}%
\typeout{** the default language instead.}%
\else
\language=\csname l@#1\endcsname
\fi
#2}}
\providecommand{\BIBdecl}{\relax}
\BIBdecl

\bibitem{george2005neural}
D.~George and J.~Hawkins, ``A hierarchical bayesian model of invariant pattern
  recognition in the visual cortex,'' in \emph{Neural Networks, 2005. IJCNN
  '05. Proceedings. 2005 IEEE International Joint Conference on}, vol.~3, July
  2005, pp. 1812--1817 vol. 3.

\bibitem{hawkins2006hierarchical}
J.~Hawkins and D.~George, ``Hierarchical temporal memory: Concepts, theory and
  terminology,'' Technical report, Numenta, Tech. Rep., 2006.

\bibitem{melis2009evaluation}
W.~J. Melis, S.~Chizuwa, and M.~Kameyama, ``Evaluation of the hierarchical
  temporal memory as soft computing platform and its vlsi architecture,'' in
  \emph{39th International Symposium on Multiple-Valued Logic}.\hskip 1em plus
  0.5em minus 0.4em\relax IEEE, 2009, pp. 233--238.

\bibitem{zyarah2015design}
A.~M. Zyarah, ``Design and analysis of a reconfigurable hierarchical temporal
  memory architecture,'' Master's thesis, 2015.

\bibitem{fan2016Hierarchical}
D.~Fan, M.~Sharad, A.~Sengupta, and K.~Roy, ``Hierarchical temporal memory
  based on spin-neurons and resistive memory for energy-efficient
  brain-inspired computing,'' \emph{IEEE Transactions on Neural Networks and
  Learning Systems}, vol.~27, no.~9, pp. 1907--1919, Sept 2016.

\bibitem{ibrayevdesign}
T.~Ibrayev, A.~P. James, C.~Merkel, and D.~Kudithipudi, ``A design of htm
  spatial pooler for face recognition using memristor-cmos hybrid circuits,''
  in \emph{2016 International Symposium on Circuits and Systems (ISCAS)}.\hskip
  1em plus 0.5em minus 0.4em\relax IEEE, 2016.

\bibitem{fedorova2016htm}
A.~P. James, I.~Fedorova, T.~Ibrayev, and D.~Kudithipudi, ``Htm spatial pooler
  with memristor crossbar circuits for sparse biometric recognition,''
  \emph{IEEE Transactions on Biomedical Circuits and Systems}, vol.~PP, no.~99,
  pp. 1--12, 2017.

\bibitem{hawkins2010hierarchical}
J.~Hawkins, S.~Ahmad, and D.~Dubinsky, ``Hierarchical temporal memory including
  htm cortical learning algorithms,'' \emph{Techical report, Numenta, Inc,
  Palto Alto http://www. numenta.
  com/htmoverview/education/HTM\_CorticalLearningAlgorithms. pdf}, 2010.

\bibitem{Farahmand2009Online}
N.~Farahmand, M.~H. Dezfoulian, H.~GhiasiRad, A.~Mokhtari, and A.~Nouri,
  ``Online temporal pattern learning,'' in \emph{2009 International Joint
  Conference on Neural Networks}, June 2009, pp. 797--802.

\bibitem{Ramli2015Pattern}
I.~Ramli and C.~Ortega-Sanchez, ``Pattern recognition using hierarchical
  concatenation,'' in \emph{Computer, Control, Informatics and its Applications
  (IC3INA), 2015 International Conference on}, Oct 2015, pp. 109--113.

\bibitem{Csapo2007Object}
A.~B. Csapo, P.~Baranyi, and D.~Tikk, ``Object categorization using
  vfa-generated nodemaps and hierarchical temporal memories,'' in
  \emph{Computational Cybernetics, 2007. ICCC 2007. IEEE International
  Conference on}, Oct 2007, pp. 257--262.

\bibitem{fyfe2007hebbian}
C.~Fyfe, \emph{Hebbian learning and negative feedback networks}.\hskip 1em plus
  0.5em minus 0.4em\relax Springer Science \& Business Media, 2007.

\bibitem{yakopcic2013memristor}
C.~Yakopcic, T.~M. Taha, G.~Subramanyam, and R.~E. Pino, ``Memristor spice
  model and crossbar simulation based on devices with nanosecond switching
  time,'' in \emph{Neural Networks (IJCNN), The 2013 International Joint
  Conference on}.\hskip 1em plus 0.5em minus 0.4em\relax IEEE, 2013, pp. 1--7.

\bibitem{biolek2016}
D.~Biolek, Z.~Kolka, V.~Biolkova, and Z.~Biolek, ``Memristor models for spice
  simulation of extremely large memristive networks,'' in \emph{2016 IEEE
  International Symposium on Circuits and Systems (ISCAS)}.\hskip 1em plus
  0.5em minus 0.4em\relax IEEE, 2016, pp. 389--392.

\bibitem{maan2016survey}
A.~K. Maan, D.~A. Jayadevi, and A.~P. James, ``A survey of memristive threshold
  logic circuits,'' \emph{IEEE Transactions on Neural Networks and Learning
  Systems}, vol.~28, no.~8, pp. 1734--1746, Aug 2017.

\bibitem{maan2015memristor}
A.~K. Maan, A.~P. James, and S.~Dimitrijev, ``Memristor pattern recogniser:
  isolated speech word recognition,'' \emph{Electronics Letters}, vol.~51,
  no.~17, pp. 1370--1372, 2015.

\bibitem{olga2017htm}
O.~Krestinskaya, T.~Ibarayev, and A.~James, ``Hierarchical temporal memory
  features with memristor logic circuits for pattern recognition,'' \emph{IEEE
  Transactions on Computer-Aided Design of Integrated Circuits and Systems},
  vol.~PP, 2017.

\bibitem{lazzaro1988winner}
J.~Lazzaro, S.~Ryckebusch, M.~A. Mahowald, and C.~A. Mead, ``Winner-take-all
  networks of o (n) complexity,'' CALIFORNIA INST OF TECH PASADENA DEPT OF
  COMPUTER SCIENCE, Tech. Rep., 1988.

\bibitem{mostafa2016process}
H.~Mostafa and Y.~Ismail, ``Process variation aware design of multi-valued
  spintronic memristor-based memory arrays,'' \emph{IEEE Transactions on
  Semiconductor Manufacturing}, vol.~29, no.~2, pp. 145--152, 2016.

\bibitem{martinez1998ar}
A.~Mart{\i}nez and R.~Benavente, ``The ar face database,'' \emph{Rapport
  technique}, vol.~24, 1998.

\end{thebibliography}

%\begin{figure}[t]
%\caption{Figure caption.}\label{f1}
%\end{figure}

%\begin{table*}
%\caption{} \label{t1}
%\begin{tabular}{lll}
%\hline
%&&\\
%&&\\
%\hline
%\end{tabular}
%\end{table*}

%%%%%%%%%%% The bibliography starts:
% \begin{thebibliography}{9}

% \bibitem{r1}

% \bibitem{r2}

% \end{thebibliography}

\end{document}